\documentclass[usenatbib,useAMS]{mn2e}

\usepackage{graphicx}

\usepackage{aas_macros}

\title[]{Mass segregation in the young open cluster NGC 2547}

\author[]{S.\,P.\ Littlefair$^{1}$, Tim Naylor$^{1}$, R.\,D.\ Jeffries$^{2}$, C.\,R.\,Devey${^2}$, S.\,Vine${^3}$\\
$^1$School of Physics, University of Exeter, Stocker Road,
Exeter EX4 4QL, UK \\
$^2$Department of Physics, Keele University, 
Staffordshire, ST5 5BG, UK \\
$^3$Department of Physics and Astronomy, St Andrews University \\}

\date{\center{\Large Submitted for publication in the Monthly Notices of the
Royal Astronomical Society \\ 
\vspace{.5cm} \today}} 

\begin{document}
\maketitle

\begin{abstract} 
  We present a study of mass segregation of the young (20--35 Myr
  isochronal age), open cluster NGC 2547.  We find good evidence that
  mass segregation exists in NGC 2547 down to 3M$_{\odot}$, and weak
  evidence for mass segregation down to 1M$_{\odot}$. Theoretical
  models of an initially unsegregated model of NGC 2547 using the {\sc
    NBODY2} code show weaker mass segregation, implying that at least
  some of the observed mass segregation has a primordial origin.  We
  also report the discovery of three possible escaped cluster members,
  which share the proper motion and colours of the cluster, but lie
  nearly a degree from the cluster centre.
  
\end{abstract} 


\section{Introduction}
\label{sec:introduction}
Most stars do not form in isolation, but in rich clusters of stars,
with sizes ranging from several tens of stars to clusters containing
several thousand stars. Observations of young clusters typically find
that the most massive stars are located near the cluster centre
\citep{hillenbrand98,raboud99}. What is not certain is whether this
reflects the initial conditions of star formation, or the process of
dynamical evolution.

Dynamical evolution in a stellar cluster drives the system towards
equipartition, the natural result of this being that the lower-mass
stars attain higher velocities and hence occupy larger orbits around
the cluster centre. In turn, the higher mass stars will sink towards
the cluster centre. \cite{bonnell98} showed that the timescale for
significant dynamical mass segregation to occur is comparable with the
relaxation time for a cluster,
\begin{equation}
\tau_r \approx \frac{N}{8\ln N} \tau_{cr}, 
\label{eq:relax}
\end{equation}
where $N$ is the number of stars within the cluster and $\tau_{cr}$
is the crossing time.  Another measure of the clusters relaxation
timescale is its half-mass, or median timescale \citep{degrijs02},
\begin{equation}
\tau_{r,h} = \left( 8.92\times10^5 \right) \frac{M_{tot}^{1/2}}{\langle m \rangle}
\frac{R_h^{3/2}}{\log_{10}\left( 0.4 M_{tot}/ \langle m \rangle \right)}\, {\rm yr},
\end{equation}
where $R_h$ is the half-mass radius in parsecs, $\langle m \rangle$ is
the mean stellar mass, and $M_{tot}$ is the total cluster mass, both
expressed in solar units. This equation represents the relaxation timescale
for the inner half of the cluster mass, for stars with characteristic
velocity dispersions and average masses.

In practice the situation is complicated because mass
segregation is a local, not a global effect. The mass segregation
timescale is smaller for the higher mass stars
\citep[e.g.][]{kontizas98}, is smaller in the cluster core than in the
outer regions \citep[e.g][]{hillenbrand98} and is larger for stars
with higher velocity dispersions. Even the concept of a local
relaxation time is approximate as mass segregation is an ongoing
process. For example, the relaxation time for high mass stars is a
decreasing function of time, as the orbits of the high mass stars
evolve closer to the cluster core with time. Complicating matters even
further, the time-scale for a cluster to lose all traces of its
initial conditions depends on the number of stars \citep{bonnell98}
the frequency of binary stars and the slope of the mass function.

Bearing this is mind, it is not surprising that it is often difficult
to establish how much dynamical evolution a cluster should have
undergone. Determining this is important because the {\em initial}
distribution of mass within the cluster may not be uniform, due to the
details of the star formation process.  Although in the classic
picture of star formation \citep{shu87} mass segregation does not
occur, there are competing models in which the most massive stars are
expected to form near the cluster centre. One such model proposes
that high mass stars are formed by mergers of lower mass protostellar
clumps \citep[e.g.][]{bonnell97} - implying that high mass stars
form in the centre of the cluster, where the density of these clumps
is highest. Alternatively, it has been suggested that protostars are
competing for the accretion of gas \citep{bonnell01} - implying that
the higher mass stars form in the centre of the cluster because that
is where the gas density is highest.

Studies of mass segregation in very young open clusters seem to
suggest that primordial mass segregation is a reality; mass
segregation is already present in the clusters NGC 2024, NGC 2071 and
Mon R2 \citep{lada91, hillenbrand98}, despite the fact that these
clusters are still embedded in their parental cloud and are not even a
crossing time old. Both \cite{hillenbrand98} and \cite{bonnell98}
argue convincingly that the mass segregation observed in the Orion
Nebular Cloud is, at least partly, due to primordial mass segregation.

Observations of older clusters usually reveal the mass segregation
expected of a dynamically relaxed cluster. Mass segregation has, for
example, been observed in the Pleiades, NGC 2516, Praesepe and M67
\citep{raboud98,mathieu86,jeffries01}, with ages of 100, 150, 800 and
5000 Myr respectively. However, \cite{raboud98} found the degree of
mass segregation for stars between 1.5 and 2.3 $M_{\odot}$ is less
pronounced in Praesepe than in the Pleiades, which is unexpected given
their ages.

Here we present a study of mass segregation in the young open cluster
NGC 2547. At an age of 20--35 Myr this cluster represents an
intermediary between very young open clusters like the Orion Nebular
Cluster, and older clusters like the Pleiades. Studies of mass
segregation in clusters of this intermediate age are useful because
mass segregation should just be starting to appear. Furthermore, it
has been suggested that a subset of clusters at this age should show
no mass segregation, due to violent relaxation following the loss of
the highest mass stars through supernovae \citep{raboud98}. In
section~\ref{sec:cat}, we describe the catalogue used to select
cluster members.  Section~\ref{sec:selec} outlines the process of
member selection, whilst in section~\ref{sec:masseg} we determine the
degree of mass segregation present in the cluster, and compare this to
expectations from theoretical arguments.

\subsection{NGC 2547}
\label{subsec:ngc}
The open cluster NGC2547 (= C0809-491) lies at Galactic co-ordinates
$l=264.45^\circ$ $b=-8.53^\circ$ ($\alpha$ = 08 10 25.7, $\delta$ =
-49 10 03; J2000).  \cite{claria82} derived a reddening for the
cluster of $E(B-V)=0.06$. \cite{naylor02} revisited the dataset of
\cite{jt98}, improving their photometry by using optimal photometry
techniques. They determined an age of 20--35 Myr and an intrinsic
distance modulus of 8.00--8.15 magnitudes (400--425 parsecs). They
estimated the cluster mass at 370 M$_{\odot}$. Hence, NGC 2547 is much
less massive than previously well studied open clusters like the
Pleiades and NGC 2516 and less massive than the well studied Orion
Nebular Cluster (ONC).

NGC 2547 is a cluster of intermediate age, although the exact age
remains uncertain. Whilst isochronal fitting to colour magnitude
diagrams suggests an age of 20--35 Myr \citep{naylor02}, the location
of the lithium depletion boundary suggests that the age is between 35
and 54 Myr \citep{oliveira03}. For the purposes of this paper we adopt
the isochronal age, although our results are not affected if the older
age is assumed. 

\section{The Catalogue}
\label{sec:cat}
\begin{figure}
\centering
\includegraphics[scale=0.33,angle=90]{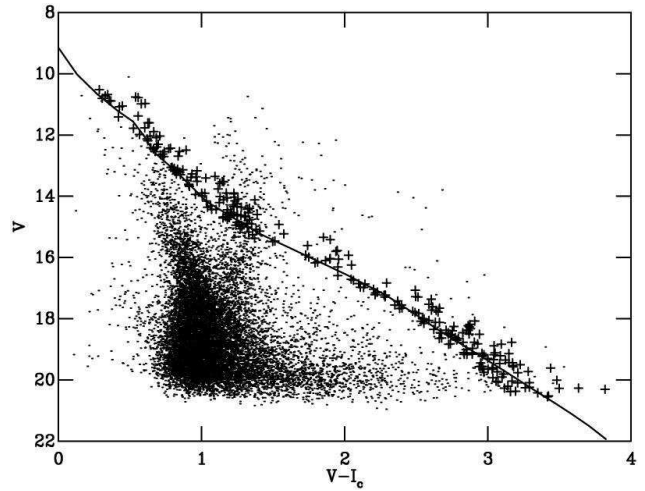}
\caption{$V$ vs.\,$V-I_{\rm c}$ for NGC 2547.
  The solid line is the best fitting D'Antona \& Mazitelli isochrone
  from \protect\cite{naylor02} and crosses represent candidate members
  selected in section~\ref{sec:selec}. Sources with a signal-to-noise
  ratio worse than 10 have been omitted.}
\label{fig:cmd}
\end{figure}

\begin{figure}
\centering
\includegraphics[scale=0.33,angle=90]{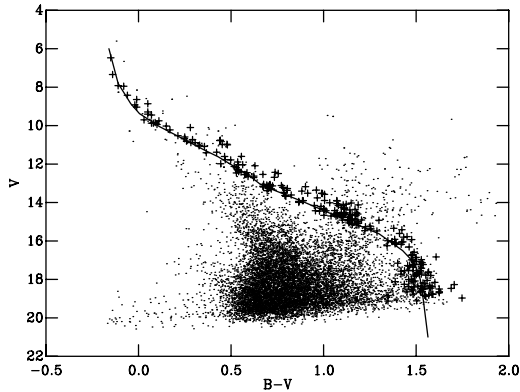}
\caption{$B$ vs.\,$B-V$ for the enhanced catalogue.
  The solid line is the best fitting D'Antona \& Mazitelli isochrone
  from \protect\cite{naylor02}. The isochrone has been extended above
  V=8 using the models of \protect\cite{schaller92}. Crosses represent
  candidate members selected in section~\ref{sec:selec}. Sources with
  a signal-to-noise ratio worse than 10 have been omitted.}
\label{fig:cmd2}
\end{figure}

The analysis which follows in sections~\ref{sec:selec} and
\ref{sec:cluster} is based upon the wide photometric catalogue of
\cite{naylor02}. The catalogue is the result of a $BVI_{\rm c}$
survey of NGC 2547, centred on $\alpha$ = 08 10 17.4, $\delta$ =
-48 57 00. The survey consists of 9
fields  covering a total of 34$\times$34 arcmins. The survey contains no
information for the brightest objects, as stars brighter than $V
\simeq 8$ saturated in the exposure times used. \cite{naylor02}
obtained this information from the photometry of \cite{claria82}. As a
consequence of this, the brightest stars only have $V$ magnitudes and
$B-V$ colours. The catalogue in which the photometry for the brightest
stars has been added is referred to as {\em enhanced}.  Colour
magnitude diagrams for the enhanced catalogue are shown in
figures~\ref{fig:cmd} and \ref{fig:cmd2}.

\subsection{Completeness}
\label{subsec:complete}
\begin{figure}
\centering
\includegraphics[scale=0.4,angle=90]{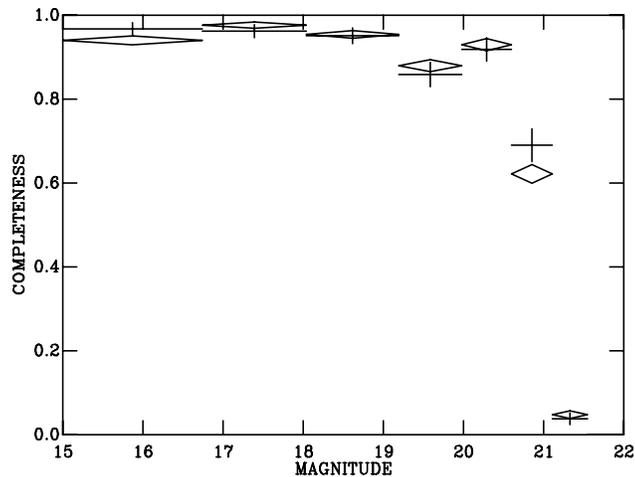}
\caption{The completeness function for the catalogue. The completeness
  function inside a radius of 10 arcmins is marked by error bars, whilst
  the completeness function outside a radius of 10 arcmins is shown using
  error boxes.  The completeness functions are remarkably similar,
  showing that the completeness of the wide catalogue does not vary
  between the inner and outer regions.}
\label{fig:complete}
\end{figure}

The catalogue of \cite{naylor02} has been produced using optimal
photometry techniques, detailed within that paper. These techniques
have the advantage of producing colour-magnitude diagrams with well
understood completeness properties. \cite{naylor02} measure the
completeness of our catalogue in the regions where it overlaps their,
deeper, catalogue, and find that it is essentially complete, down to a
$V$ magnitude of 20.5.  Also, the completeness drops sharply from 90\%
to 10\% in half a magnitude.  In this paper we are interested in the
completeness of our catalogue as a function of position. We assessed
this by injecting a simulated sequence of pre-main sequence stars into
the data and determining what fraction are selected as cluster members
in our final catalogue \citep[see][for details]{naylor02}.
Figure~\ref{fig:complete} shows that the completeness properties do not
vary between the inner and outer regions, and that the wide catalogue is
complete to $V=20.5$, with a very sharp cut-off below this. These
properties mean that corrections for completeness can be ignored in
the following analysis, without introducing significant error.

\section{Selection of cluster members}
\label{sec:selec}
Selection of candidate members of NGC 2547 proceeds largely as
described in \cite{naylor02}. Each star is tested against a number of
criteria for membership. If a star does not fail any of these tests
then it is retained as a candidate member. Briefly, these tests
consist of investigating whether a star is: (a) close to the best
fitting $V$ vs $V-I_{\rm c}$ isochrone, defined in \cite{naylor02};
(b) close to the $V$ vs $B-V$ isochrone; (c) close to the $V-I_{\rm
  c}$ vs $B-V$ locus for cluster members. At the referee's request, we
also attempted to use the UCAC1 proper motion catalogue
\citep{zacharias00} to refine our membership selection, but the data
were not precise enough for our purposes.

In addition we apply a final test for possible binarity. If a star
lies more than 0.3 mag. above the $V$ vs $V-I_{\rm c}$ isochrone (or
the $V$ vs $B-V$ isochrone for stars with $V-I_{\rm c}<0.5$) we class
them as candidate unresolved binary systems. \cite{naylor02} calculate
that this method is sensitive to binaries with mass ratio $q\geq 0.6$
for $V<14$ and $q\geq0.5$ for fainter cluster members.

For the brightest objects, for which we have no $V-I_{\rm c}$ colour,
we were forced to use slightly different selection criteria. For these
objects, membership was allocated to those stars which are close to
the $V$ vs $B-V$ isochrone.  We reject those measurements which are
affected by bad pixels, have poor sky subtraction, are flagged as
non-stellar or which have uncertainties of greater than 0.2 mag. Above
$V=8$ the $V$ vs $B-V$ isochrone lies nearly vertical and here members
have been selected ``by-eye''.  From the objects with no $V-I_{\rm c}$
colour, we selected a further 24 members, out of 39 objects.

For this paper we have introduced two further selection criteria. The
membership selection for the very brightest stars was checked using
the {\small HIPPARCHOS} proper motions \citep{baumgardt00} and the
radial velocities from the Revision of the General Catalogue of Radial
Velocities \citep{evans67}. In each case, our results for membership
using the photometric cut described above agreed with the membership
results using the proper motions and radial velocities. The only
exception was the variable star KW Vel. Although this star was
selected as a member on the basis of its photometric colours,
\cite{baumgardt00} classified the object as a non-member on the basis of
its proper motion. Hence, we removed KW Vel from our
catalogue of members. As an aside, three objects (HIP \# 40336, 40385
and 40427) were discovered in the {\small HIPPARCHOS} proper motions
of \citet{baumgardt00} which lay well outside our surveyed region, but
were selected as cluster members on the basis of their proper motions
and {\small HIPPARCHOS} photometry.  These stars lie about a degree
away from the cluster centre, and we identify them as potential
escaped cluster members. 

We used the method described above to find 327 cluster
candidates from the enhanced catalogue, of which 95 are probable
unresolved, high mass ratio binaries. Figure~\ref{fig:cmd} shows the
$V$ vs.\,$V-I_{\rm c}$ colour-magnitude diagram, with cluster
candidates indicated.

\subsection{Field Star Contamination}
\label{subsec:contam}
We are confident that the membership selection criteria have included
almost all the true cluster members. A crucial question regards the
number of non-members which we have erroneously classed as members. We
note that the NGC 2547 isochrone for PMS stars lies more than a
magnitude above the ZAMS in the colour-magnitude diagram, and hence
clear of background contamination. The exception is between $14.0 < V
< 15.5$ (0.75--0.9M$_{\odot}$) where a ``finger'' of background giants
intrudes onto the PMS locus. Here, integration of an interpolation of
the density of stars above and below the cluster sequence suggests
that the field star contamination could be as high as 40\% in this
region of colour-magnitude space.

The assumption that background contamination is negligible outside the
contaminating finger has been tested by \cite{jeffries00}, who
performed high-resolution spectroscopy of 23 of the stars selected as
cluster members by \cite{naylor02}, with magnitudes in the range $12 <
V < 15$. From the radial velocities there is good evidence that at
least 20, and probably all, of these stars are indeed cluster members.
Therefore, following \cite{naylor02}, we assume contamination is
negligible in the analysis which follows.

\section{NGC 2547 as a cluster}
\label{sec:cluster}

\subsection{The cluster centre}
\label{subsec:centre}

There are several ways to define a cluster centre; theoretically, it
can be defined as the centre of mass, or the location of the deepest
part of the gravitational potential, observationally the centre is
often defined as the region of highest surface brightness, or the
region containing the largest number of objects. Here we define the
cluster centre as the being the location of maximum stellar density.

\begin{figure*}
\centering
\includegraphics[scale=0.7,angle=-90]{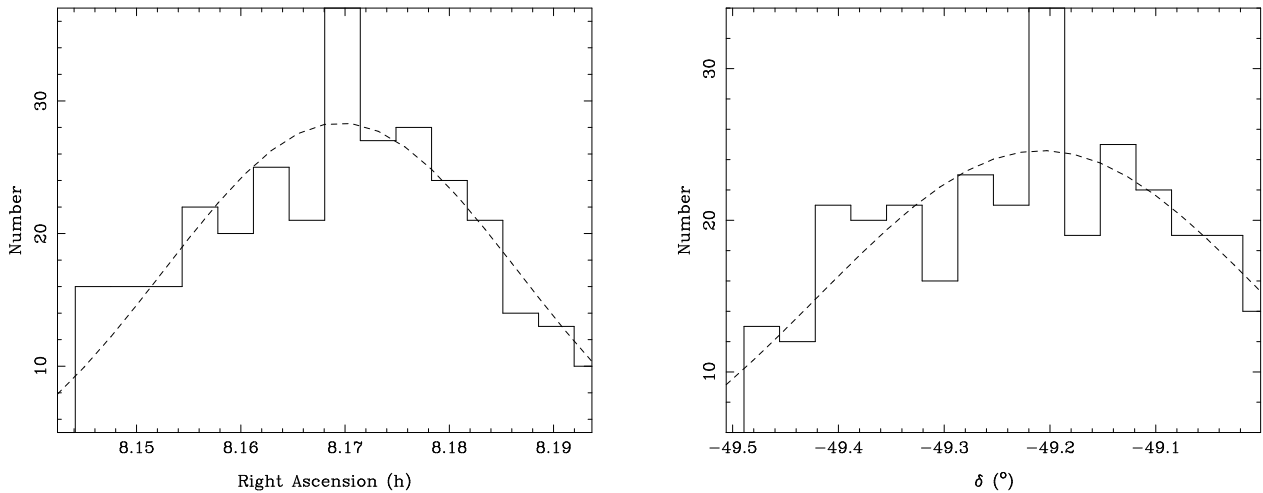}
\caption{Profiles of stellar counts across NGC 2547. Only
candidate members of NGC 2547 were included in the census. 
The Gaussian fits used to determine the cluster centre are
shown as dashed lines.}
\label{fig:prof}
\end{figure*}

The cluster centre was found by fitting Gaussians to the profiles of
star counts in right ascension and declination. These profiles are
shown in figure~\ref{fig:prof}.
This method gives both a location for the cluster centre, and a formal
error. Using this method, the cluster centre is found to be at $\alpha
= 8.169^{+0.02}_{-0.01}$ hours and $\delta = -49.20^{+0.02}_{-0.02}$
degrees.

\subsection{Is NGC 2547 bound?}
\label{subsec:bound}
We can determine an approximation to (strictly, a lower limit to) the
half-mass radius of the cluster, by finding the radius which contains
half the total mass in our survey.  Unresolved binarity can be a
problem for this technique if the binary fraction or distribution of
mass ratios is a strong function of position within the cluster.
Hence, we determined the half-mass radius by two methods; first by
ignoring the problems raised by unresolved binary companions, and
second using the scheme outlined in \cite{naylor02} to account for the
high mass ratio binaries.

We find that the half mass radius of NGC 2547 is 10.7 arcmins,
regardless of whether binaries are accounted for or not. We note here
that the method used assumes spherical symmetry for the cluster. The
effect of this assumption on the value of the half-mass radius was
investigated by determining the half mass radius obtained for four
quadrants of the cluster. The half mass radii found for these four quadrants
ranged from 10 to 11 arcmins. Hence, we believe that the assumption of
spherical symmetry does not introduce errors greater than 10\%.

Using the half-mass radius as a lower limit to the total radius of the
cluster, and the the velocity dispersion of \cite{jeffries00} of
$\bar{v}\le1$\,km\,s$^{-1}$, the virial mass is $M \simeq 2
R_{tot}\bar{v}^2/G \approx 600M_{\odot}$. This value is highly
uncertain, however, because the total radius of the cluster, and the
velocity dispersion are basically unknown. The final result could
easily vary by a factor of two either way.  The total mass of NGC 2547
is 370$M_{\odot}$, and therefore it is not possible to determine if
NGC 2547 is bound with the data available at present. However, given
its continued existence after 20--35 Myr, it seems likely that the
cluster is at least close to being bound.

\section{Mass segregation in NGC 2547}
\label{sec:masseg}
\begin{figure}
  \centering \includegraphics[scale=0.4,angle=90]{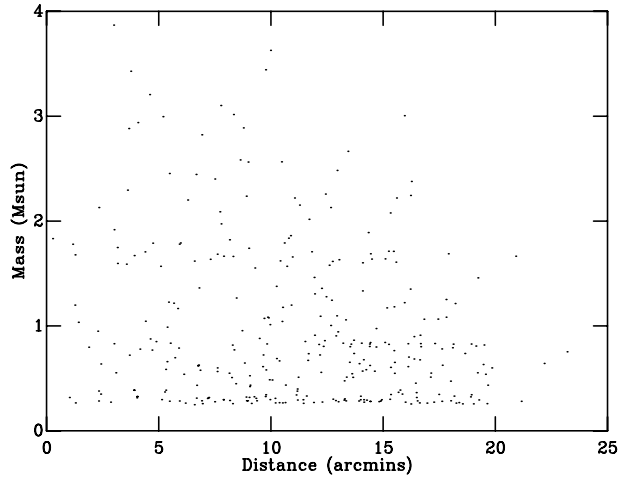}
\caption{Mass segregation in NGC 2547}
\label{fig:vvr}
\end{figure}

The evidence for mass segregation in NGC 2547 can be seen from a plot
of radius against mass for the cluster members (figure~\ref{fig:vvr}).
Masses were allocated to stars following the procedure in
\cite{naylor02}, which makes corrections for binarity.  The key
features of this plot are: the ``finger of contamination'', see
section~\ref{subsec:contam}, which is just visible as a strip of
increased stellar density between 10 and 20 arcmins and
$0.8<M/M_{\odot}<0.9$, and the sharp drop in the number of stars at
around 18 arcmins. The mass segregation is visible as an absence of
stars in th top-right of the diagram. In particular, the maximum
radius seems to decrease above 2M$_{\odot}$.

In an attempt to quantify the significance of this feature, we start
from the null hypothesis that there is no mass segregation in
NGC~2547. The approach used to test this hypothesis is described
below.
\begin{itemize}
\item{We use the radial distribution and mass function for NGC 2547 to
    create a model of the null hypothesis, in which there is no mass
    segregation (i.e. the radial distribution is independent of
    mass).}
\item{The distribution of stars in NGC 2547 in the V-magnitude, radius
    plane is compared to that of the model, by measuring the
    D-statistic\footnote{The D statistic is the largest difference
      between the fractions of data points which lie in the four
      quadrants around a point ($x$,$y$), and the corresponding
      fractions of data points from a model which corresponds to the
      null hypothesis, where the point ($x$, $y$) has been selected to
      maximise the D-statistic} from a two-dimensional, 1-sample K-S
    test.}
\item{We now generate many fake clusters, which share the mass
    function and radial distribution of NGC 2547, but show no mass
    segregation.}
\item{We now perform the same K-S test applied previously to NGC 2547
    to our fake clusters, in order to discover the distribution of the
    D-statistic in the null hypothesis.}
\item{The D statistic from the real NGC 2547 data is compared to the
    distribution in the null hypothesis}
\end{itemize}
  
The test described above gives a formal probability that there is {\bf
  no} mass segregation in NGC2547 (mass function is independent of
radius) of 95 per cent, in stark contrast to what we would expect from
visually inspecting figure~\ref{fig:vvr}. Hence we see that the 2-d
K-S test is not a powerful test for the presence of mass segregation.
This lack of power arises because there are many possible deviations
from the null hypothesis (independent mass and radius distributions)
which do not correspond to the distributions expected from mass
segregation. In testing the distribution of stars in NGC 2547 against
all these possible deviations, we lose the sensitivity to detect the
deviation we are interested in. We can see this is true if we
formulate a different test, which only looks for deviations from the
null hypothesis in the sense {\em expected} from mass segregation. In
this test we generated 8000 fake clusters, of the same size as NGC
2547, which show no mass segregation. We then asked how many of these
clusters showed a lack of high mass stars at large radii which was
comparable with the data. Strictly, we determined the percentage of
our clusters in which there was fewer than two stars whose radii in
arcminutes obeyed $R \ge 30-5M$, where M is the stellar mass in solar
units (NGC 2547 has only one such star). Only $\sim 2$\% of our fake
clusters met our criteria.  What this test shows is that it is
possible to gain sensitivity to the presence of mass segregation, by
only testing for deviations in the sense expected. Hence, increased
power comes at the cost of increased bias.

\begin{figure}
\centering
\includegraphics[scale=0.3,angle=-90]{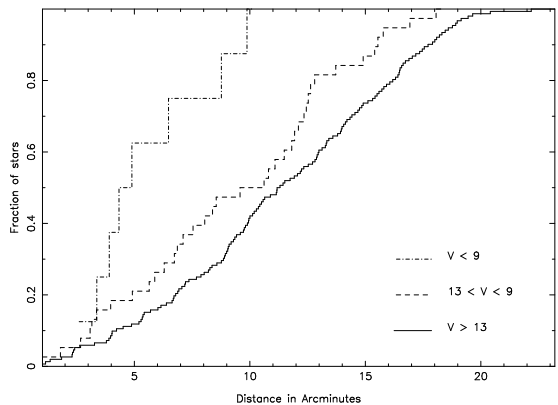}
\caption{Cumulative fraction profiles of distance from cluster centre for stars
  of  different mass  in NGC  2547. The  high mass  stars  ($V<9$) are
  significantly centrally condensed.  The distribution of intermediate
  mass stars  ($9<V<13$, corresponding to  $1<M<3$ $M_{\odot}$) appear
  more centrally condensed than stars  of low mass $V>13$, but this is
  not formally significant (see text for details).}
\label{fig:cum_frac}
\end{figure}
Having determined the presence of mass segregation in NGC 2547, it is
pertinent to ask at what masses it is present.
Figure~\ref{fig:cum_frac} shows cumulative radial distributions for
non-binary members of NGC 2547 in 3 mass bins. The ``finger'' of
background giants (i.e. stars in the $V$ magnitude range 14--15.5) has
been excluded. A 1-D K-S test gives a 0.5 per cent probability that
the $V<9$ (approximating to $M>3$ $M_{\odot}$) stars have the same
radial distribution as the stars with $V>9$, and a 3 per cent
probability that the $V<9$ stars have the same distribution as the
stars with $9<V<13$.  The results described above do not change if we
move the cluster centre around 1$\sigma$ in RA and Dec (i.e a 1.4
$\sigma$ shift in centre position).  In summary, the data show good
evidence that the stars above 3$M_{\odot}$ exhibit mass segregation.

A word of caution: the numbers above should not be regarded as
representative of the formal significance, because we chose the mass
bins as such because there were no stars with $V<9$ outside the half
mass radius (i.e. we have seen what we think is an effect in the data,
and then set out to prove it is present). The K-S test assumes that
the bin boundaries are chosen at random and will therefore have
overestimated the significance of our results. The situation is
somewhat analogous to the difficulties we found in applying the 2-d
K-S test - the ability to detect mass segregation has come at the cost
of some unknown level of bias. We also note here that, to our
knowledge, all previous studies of mass segregation have used a
similar 1-d K-S test, and suffer from the same limitations.

If there is mass segregation above 3M$_{\odot}$, and none in the
lowest mass stars in our census, it is relevant to ask at what point
mass segregation becomes important.  This is revealed by the results
of the 1-d KS test between the radial distributions of stars with
$9<V<13$ and $V>13$ ($V=13$ is approximately 1$M_{\odot}$).  The 1-d
K-S test gives a probability that these distributions are the same of
7 per cent. Although this result is not formally significant, it
suggests that some mass segregation in NGC 2547 has occurred for stars
with masses greater than somewhere between 1 and 3M$_{\odot}$.

Hence we conclude that there is good evidence for mass segregation for
the stars above 3$M_{\odot}$. Weak evidence exists that mass
segregation may have occurred down to 1$M_{\odot}$ and there is no
evidence that mass segregation has occurred for stars less massive
than this. 

\subsection{Binary Stars}

Several theoretical studies of mass segregation conclude that binary
stars tend to be more centrally condensed than single stars because
they are, on average, more massive. An early study by \cite{abt80}
found that this was indeed the case, but only for clusters older than
$10^8$\,yr. \cite{raboud94} studied the distribution of red giant
stars in 14 clusters with ages between 3$\times 10^8$ and 40$\times
10^8$\,yr. They found that only M67 (the oldest cluster observed)
showed a significant difference between the spatial distribution of
red giant binaries and single red giants, confirming an earlier result
by \citep{mathieu86}. In contrast to these results, \cite{raboud98}
find that the binary stars in the $\sim 3$\,Myr old cluster NGC 6231
are centrally condensed.

In agreement with the findings of \cite{abt80}, there is no evidence
that the binary stars in NGC 2547 are more centrally condensed than
the single stars: the 1-d K-S test gives $P_{null}=70$ per cent.

\subsection{Timescales}

The radial velocity dispersion in NGC 2547 is 1 km\,s$^{-1}$
\citep{jeffries00}, though this is dominated by errors, and should be
considered an upper limit. The total mass is 370 $M_{\odot}$
\citep{naylor02}, with 323 members down to a mass of 0.26 $M_{\odot}$
and out to radii of 20 arcmins (this paper).

With an age estimate for NGC 2547 of 20--35 Myr, this allows us to
calculate a crossing time for the cluster of $\tau_{cr} \geq 3$\,Myr.
\cite{bonnell98} show that mass segregation occurs for the higher mass
stars in a cluster on the order of the relaxation time (see
equation~\ref{eq:relax}). For NGC 2547, this gives a relaxation time of
$\tau_r \approx 20$\,Myr, comparable with the age of the cluster.

Hence, NGC 2547 lies in an interesting region of parameter space where
the cluster age is comparable to the time taken for mass segregation
to appear. Because of this, detailed modelling is necessary to
determine if the mass segregation observed in NGC 2547 is caused by
dynamical evolution, or primordial effects.

\subsection{Modelling}
\label{subsec:modelling}
\begin{figure}
\centering
\includegraphics[scale=0.3,angle=-90]{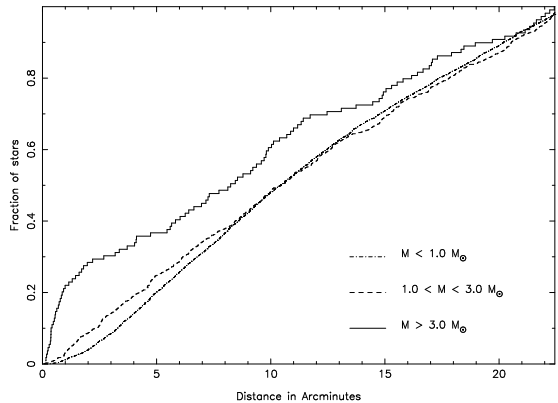}
\caption{Average cumulative fraction profiles of distance from cluster centre 
  for ten models of NGC 2547 at an age of 35 Myr. Some segregation of
  the high mass stars has begun, but mass segregation in our model is
  weaker than that observed in NGC 2547 itself.}
\label{fig:cum_frac_theory}
\end{figure}
We modelled the dynamical evolution of NGC 2547 using the {\sc nbody2}
code.  A model cluster was created by generating stars until the model
cluster matched the total mass of NGC 2547. The probability of
selecting a star of given mass and radius from cluster centre was
weighted in order to match the observed mass function and half mass
radius.  The initial conditions in the model follow a Plummer distribution
with an isothermal velocity distribution. The model contained no initial
mass segregation.
 
Figure~\ref{fig:cum_frac_theory} shows the cumulative fraction
profiles of distance from cluster centre for 10 models of NGC 2547 at
35 Myr with identical statistical properties but different initial
random realisations. It can be seen that the high mass (M $> 3$M$_{\odot}$)
stars are slightly centrally condensed, but that the low and medium mass
stars show no mass segregation. 

The model cluster shows less mass segregation at 35 Myr than observed
in NGC 2547. This is most strongly seen in the high mass stars. In the
model cluster 55\% of the high mass stars lie outside the half-mass
radius, whereas in NGC 2547 all of the high mass stars lie within this
radius. If we assume that the high mass stars in NGC 2547 have the
same radial distribution as seen in our model cluster, there is only a
0.2\% chance of finding all eight high mass stars in NGC 2547 within
the half mass radius. Hence dynamical evolution alone cannot explain
the central concentration of high mass stars in NGC 2547 and we
conclude that some, but not all of the mass segregation observed in
NGC 2547 is primordial in origin.

\section{Conclusions}
\label{sec:conclusions}

We find evidence for the onset of mass segregation amongst the high
mass ($> 3 M_{\odot}$) members of NGC 2547. Weaker evidence exists for
mass segregation down to $1 M_{\odot}$. Modelling of NGC 2547 produces
clusters with weaker mass segregation, implying a primordial origin
for some of the mass segregation observed. Hence we conclude that, in
some cases, primordial mass segregation can still be observed in
intermediate (20--35 Myr) age clusters.

It is likely that NGC 2547 has suffered the loss of high mass stars
through supernovae in its history. Supernovae are expected to have
occurred at the age of NGC 2547 for stars more massive than roughly
8M$_{\odot}$. Extrapolating the mass function of \cite{naylor02} we
would expect to find around 2 stars more massive than this in NGC
2547, implying that at least one supernova has occurred in NGC 2547 in
the past.  It is therefore difficult to reconcile the mass segregation
clearly observed in NGC 2547 with the suggestion by \cite{raboud98}
that supernovae might remove traces of primordial mass segregation
from a cluster.

\section*{\sc Acknowledgements}
We are grateful to PPARC for supporting SL. We would like to thank the
director and staff of the Cerro Tololo Interamerican Observatory,
operated by the Association of Universities for Research in Astronomy,
Inc.\,,under contract to the US National Science Foundation.

\bibliographystyle{mn2e}
\bibliography{abbrev,refs}

\begin{thebibliography}{}

\bibitem[\protect\citeauthoryear{{Abt}}{{Abt}}{1980}]{abt80}
{Abt} H.~A.,  1980, \apj, 241, 275

\bibitem[\protect\citeauthoryear{{Baumgardt}, {Dettbarn} \&
  {Wielen}}{{Baumgardt} et~al.}{2000}]{baumgardt00}
{Baumgardt} H.,  {Dettbarn} C.,    {Wielen} R.,  2000, \aaps, 146, 251

\bibitem[\protect\citeauthoryear{{Bonnell}, {Bate}, {Clarke} \&
  {Pringle}}{{Bonnell} et~al.}{1997}]{bonnell97}
{Bonnell} I.~A.,  {Bate} M.~R.,  {Clarke} C.~J.,    {Pringle} J.~E.,  1997,
  \mnras, 285, 201

\bibitem[\protect\citeauthoryear{{Bonnell}, {Bate}, {Clarke} \&
  {Pringle}}{{Bonnell} et~al.}{2001}]{bonnell01}
{Bonnell} I.~A.,  {Bate} M.~R.,  {Clarke} C.~J.,    {Pringle} J.~E.,  2001,
  \mnras, 323, 785

\bibitem[\protect\citeauthoryear{Bonnell \& Davies}{Bonnell \&
  Davies}{1998}]{bonnell98}
Bonnell I.~A.,  Davies M.~B.,  1998, MNRAS, 295, 691

\bibitem[\protect\citeauthoryear{Clari\'a}{Clari\'a}{1982}]{claria82}
Clari\'a J.,  1982, A\&ASS, 47, 323

\bibitem[\protect\citeauthoryear{{de Grijs}, {Gilmore}, {Johnson} \&
  {Mackey}}{{de Grijs} et~al.}{2002}]{degrijs02}
{de Grijs} R.,  {Gilmore} G.~F.,  {Johnson} R.~A.,    {Mackey} A.~D.,  2002,
  \mnras, 331, 245

\bibitem[\protect\citeauthoryear{{Evans}}{{Evans}}{1967}]{evans67}
{Evans} D.~S.,  1967, in IAU Symp. 30: Determination of Radial Velocities and
  their Applications Vol.~30, {The Revision of the General Catalogue of Radial
  Velocities}.
pp~57+

\bibitem[\protect\citeauthoryear{{Hillenbrand} \& {Hartmann}}{{Hillenbrand} \&
  {Hartmann}}{1998}]{hillenbrand98}
{Hillenbrand} L.~A.,  {Hartmann} L.~W.,  1998, \apj, 492, 540+

\bibitem[\protect\citeauthoryear{Jeffries \& Tolley}{Jeffries \&
  Tolley}{1998}]{jt98}
Jeffries R.,  Tolley A.,  1998, MNRAS, 300, 331

\bibitem[\protect\citeauthoryear{{Jeffries}, {Thurston} \& {Hambly}}{{Jeffries}
  et~al.}{2001}]{jeffries01}
{Jeffries} R.~D.,  {Thurston} M.~R.,    {Hambly} N.~C.,  2001, \aap, 375, 863

\bibitem[\protect\citeauthoryear{{Jeffries}, {Totten} \& {James}}{{Jeffries}
  et~al.}{2000}]{jeffries00}
{Jeffries} R.~D.,  {Totten} E.~J.,    {James} D.~J.,  2000, \mnras, 316, 950

\bibitem[\protect\citeauthoryear{{Kontizas}, {Hatzidimitriou},
  {Bellas-Velidis}, {Gouliermis}, {Kontizas} \& {Cannon}}{{Kontizas}
  et~al.}{1998}]{kontizas98}
{Kontizas} M.,  {Hatzidimitriou} D.,  {Bellas-Velidis} I.,  {Gouliermis} D.,
  {Kontizas} E.,    {Cannon} R.~D.,  1998, \aap, 336, 503

\bibitem[\protect\citeauthoryear{{Lada} \& {Lada}}{{Lada} \&
  {Lada}}{1991}]{lada91}
{Lada} C.~J.,  {Lada} E.~A.,  1991, in ASP Conf. Ser. 13: The Formation and
  Evolution of Star Clusters {The nature, origin and evolution of embedded star
  clusters}.
pp 3--48676

\bibitem[\protect\citeauthoryear{{Mathieu} \& {Latham}}{{Mathieu} \&
  {Latham}}{1986}]{mathieu86}
{Mathieu} R.~D.,  {Latham} D.~W.,  1986, \aj, 92, 1364

\bibitem[\protect\citeauthoryear{Naylor, Totten, Jeffries, Pozzo, Devey \&
  Thompson}{Naylor et~al.}{2002}]{naylor02}
Naylor T.,  Totten E.,  Jeffries R.,  Pozzo M.,  Devey C.,    Thompson S.,
  2002, MNRAS, 335, 291

\bibitem[\protect\citeauthoryear{Olivera, Jeffries, Devey, Barrado~y
  Navascu\'{e}s, Naylor, Stauffer \& Totten}{Olivera et~al.}{2003}]{oliveira03}
Olivera J.,  Jeffries R.,  Devey C.,  Barrado~y Navascu\'{e}s D.,  Naylor T.,
  Stauffer J.,    Totten E.,  2003, astro-ph, 0303083

\bibitem[\protect\citeauthoryear{{Raboud}}{{Raboud}}{1999}]{raboud99}
{Raboud} D.,  1999, in Revista Mexicana de Astronomia y Astrofisica Conference
  Series Vol.~8, {Mass segregation in very young open clusters.}.
pp 107--110

\bibitem[\protect\citeauthoryear{{Raboud} \& {Mermilliod}}{{Raboud} \&
  {Mermilliod}}{1994}]{raboud94}
{Raboud} D.,  {Mermilliod} J.-C.,  1994, \aap, 289, 121

\bibitem[\protect\citeauthoryear{{Raboud} \& {Mermilliod}}{{Raboud} \&
  {Mermilliod}}{1998}]{raboud98}
{Raboud} D.,  {Mermilliod} J.-C.,  1998, \aap, 333, 897

\bibitem[\protect\citeauthoryear{{Schaller}, {Schaerer}, {Meynet} \&
  {Maeder}}{{Schaller} et~al.}{1992}]{schaller92}
{Schaller} G.,  {Schaerer} D.,  {Meynet} G.,    {Maeder} A.,  1992, \aaps, 96,
  269

\bibitem[\protect\citeauthoryear{{Shu}, {Adams} \& {Lizano}}{{Shu}
  et~al.}{1987}]{shu87}
{Shu} F.~H.,  {Adams} F.~C.,    {Lizano} S.,  1987, \araa, 25, 23

\bibitem[\protect\citeauthoryear{{Zacharias}, {Urban}, {Zacharias}, {Hall},
  {Wycoff}, {Rafferty}, {Germain}, {Holdenried}, {Pohlman}, {Gauss}, {Monet} \&
  {Winter}}{{Zacharias} et~al.}{2000}]{zacharias00}
{Zacharias} N.,  {Urban} S.~E.,  {Zacharias} M.~I.,  {Hall} D.~M.,  {Wycoff}
  G.~L.,  {Rafferty} T.~J.,  {Germain} M.~E.,  {Holdenried} E.~R.,  {Pohlman}
  J.~W.,  {Gauss} F.~S.,  {Monet} D.~G.,    {Winter} L.,  2000, \aj, 120, 2131

\end{thebibliography}

\end{document}